# An equitable and effective approach to introductory mechanics

Eric Burkholder[1], Shima Salehi[2], Sarah Sackeyfio[3], Nicel Mohamed-Hinds[4], Carl Wieman[2,3]

1 Department of Physics, Auburn University Auburn AL 36849
2 Graduate School of Education, Stanford University, Stanford CA 94305
3 Department of Physics, Stanford University, Stanford CA 94305
4 Department of Physics, University of Washington, Seattle WA 98105

**Abstract:**

Introductory mechanics ("physics 1") is a critical gateway course for students desiring to pursue a STEM career. A major challenge with this course is that there is a large spread in the students' incoming physics preparation, and this level of preparation is strongly predictive of a students' performance. The level of incoming preparation is also largely determined by a student's educational privilege, and so this course can amplify inequities in K-12 education and provide a barrier to a STEM career for students from marginalized groups. Here, we present a novel introductory course design to address such equity challenges in physics 1. We designed the course based on the concept of deliberate practice to give students targeted, scaffolded, and repeated opportunities to engage in research-identified practices and decisions required for effective problem-solving. We used real-world problems, as they carry less resemblance to physics high school problems, and so even the students with the best high school physics instruction have little experience or skill in solving them. The students learned the physics content knowledge they needed in future courses, particularly in engineering, and their problem-solving skills improved substantially. Furthermore, the success in the course was not correlated with incoming physics preparation, in stark contrast to the outcomes from traditional physics 1 courses. These findings suggest that we made physics 1 more equitable by employing a deliberate practice approach in the context of real-world problem-solving.

**Introduction:**

Introductory physics ("Physics 1") is typically the largest course taught by physics departments, and it has a drastic impact on students' college pathways, as it serves as a gateway/barrier to most STEM careers. It also poses a significant teaching challenge, because students come to college with such a large range of preparation, typically having anywhere from 0 to 2 years of physics covering much of the physics 1 material. Performance in Physics 1 has been reported to be strongly correlated with incoming preparation of the students[1], thus reflecting the inequities in high school physics teaching. We spent several years testing different ways to help students at our institution who had received less preparation succeed in physics 1. This included various forms of supplemental instruction and support, as well as changes in teaching methods, but the correlation between incoming preparation and course success remained stubbornly the same. Students with the lowest physics preparation continued to have a high probability of failing. This is unacceptable, as it shows that under-privileged students who made their way to college despite many societal challenges are not provided the same opportunity to thrive in STEM fields. When we talked to faculty from other institutions they reported similar experiences, suggesting this is a widespread pattern due to the strong overlap between the physics covered in a good high school physics class and the college physics 1 class.

After years of failure at addressing the preparation gap in the traditional physics 1 course, in desperation, we created an entirely new version of "Physics 1". The design philosophy of the course was driven by the data-based conclusion that it was simply not possible to compensate in the 10 weeks of physics 1 for a difference of 1-2 years of prior coverage of physics 1 material, via supplementary instruction. We set out to design a new course that would better prepare students for subsequent STEM courses, while being less dependent on high school physics experience. Hence, success in the course would not be mainly determined by physics incoming preparation, and the course would not simply amplify inequities in K-12 education. The results show that the course has succeeded in allowing students to be successful regardless of their incoming physics preparation, while leaving them prepared for subsequent courses that call upon the Physics 1 material. This article describes the course design process and its impact on creating a more equitable physics 1 course.

The course design involved three stages. *First,* we identified what content was covered in physics 1 that was of little importance in subsequent courses students would likely be taking, and so could be reduced or dropped to free up time in the schedule. As discussed below, we found several topics that are traditionally covered in physics 1 but are seldom needed in the engineering (or physics) curriculum. *Second,* and most important, we figured out how to teach essential topics with less overlap with what students had been taught in high school physics and hence, with less sensitivity to the quality of high school preparation. *Third,* we found unique new educational value, problem-solving skills, that we believed we could add that was not in the traditional physics 1 course. Despite the stated emphasis of problem-solving in a traditional physics 1 course, the students are rarely provided the opportunity to explicitly learn and practice these cognitive skills. The kinds of problems students solve, and the reasoning required, has little overlap with the real-world problem solving a student will be called to do in any career. We sought to enhance this course by improving problem solving skills using the results reported in Ref. 2, where they identified a specific set of practices and associated cognitive decisions that were used in effective problem solving in science and engineering.

Our course design combined those identified practices with the concept of deliberate practice from cognitive psychology[3] and has been proven effective in gaining mastery in many different areas, including physics at a variety of levels[4-5]. Ericsson et al.[3], introduced Deliberate practice as the essential approach for gaining expertise in nearly all subjects, and it is now widely applied. Deliberate practice for learning a new skill can be summarized by the following steps:

1. That skill (learning goal) should be divided into specific, simpler sub-skills (sub learning goals)
2. Corresponding practice/learning activities should be developed for each sub-skill, appropriate for the learners' level of incoming skill, challenging but attainable with effort.
3. Learners should complete these activities and receive timely guiding feedback on their performance.
4. Learners should be provided the opportunity to repeat the practice to incorporate the feedback and improve their performance in the given sub-skill, and then in the combination of sub-skills.

In the context of physics 1, teaching problem-solving involves having the students repeatedly engaging in activities targeting specific problem-solving sub-skills in the context of a suitably challenging "real-world" (not artificially constrained) problem. While they do this, they receive guiding feedback on the quality of their performance and how to improve.

We designed problems appropriate for the students incoming level of preparation and becoming increasingly more difficult as the term progressed. The problems required most of the basic physics 1 material (i. e. Newton's laws, statics, some dynamics, and energy forms and conservation), but these were more realistic than essentially all textbook and exam problems as illustrated in table 1. They were like "context rich problems" [6] but with additional constraints on the design. The problems also had to pass the test "Would someone other than a physics teacher care about the answer to this?"

| Textbook Problem[7] | Authentic Problem |
|---|---|
| You are driving to the grocery store at 20 m/s. You are 110 m from an intersection when the traffic light turns red. Assume that your reaction is 0.50s and that your car brakes with constant acceleration. What magnitude braking acceleration will bring you to stop exactly at the intersection? | You are an engineer designing a pinball machine. Determine the necessary spring constant of the spring needed to launch a typical pinball to the top of the game. |

**Table 1: An example of the contrast between textbook[7] and real-world problems. The textbook problem provides all of the information needed to solve the problem, specifies the assumptions to be made (i.e., constant acceleration), and has one correct solution path. The authentic problem has multiple potential solution paths, involves reflecting on prior knowledge, planning the solution path, deciding on what information is needed, finding that information, and making and justifying assumptions.**

While there has been extensive previous work focusing on teaching problem-solving to students[8-10], the problems students used in these works were mostly standard physics textbook problems. These require few if any of the decisions requited to solve real-world problems, and so solving these problems does not necessarily develop mastery in solving real-world problems.

Designing the course around real-world problem-solving has important implications for equity. Because few if any students had previously encountered real-world physics problems, students with extensive high school physics preparation did not start with a great advantage. While these students knew basic formulae and concepts, they did not know how to apply those ideas to solve real physics problems. Our hypothesis (later confirmed) was that the strong dependence of course performance on prior preparation would disappear when students were no longer using physics in artificial contexts. Thus, the incorporation of real world problem-solving both leaves students better prepared to use physics in their careers and (as we discuss later) eliminates the gap in course performance between students with and without prior physics preparation. We also hoped that seeing how physics could be used to solve meaningful real-world problems would give students a greater appreciation of the value of physics and its broader application.

**Problem-Solving Template:**

During the course, the students followed a template to guide them through the problem-solving decision steps. A variety of problem-solving templates for use in physics have been presented[7-10]. We used a version based on the decisions reported in Ref. 2, adapted to the material of this course. The template explicitly engaged students in different problem-solving sub-skills and allowed the instructors and TAs to give targeted grading and feedback on student performance on each of the sub-skills. Students repeatedly solved problems following this template throughout the course, so they had the opportunity to incorporate feedback from previous rounds to improve their performance in the various essential problem-solving sub-skills.

An outline of the template and its relationship to problem-solving practices and decisions is shown in Table 2. This template included: identifying the key features; deciding what information was required to solve the problem and how to find that information; deciding on and writing down a plan for the solution process; deciding on appropriate approximations; and deciding if final solution makes sense.

The template used the concept of "structured autonomy," where students were provided with structure showing what they had to decide/do, but not how to do it. This template-based problem-solving approach was then incorporated into all course elements: small group active-learning activities in class; homework; and exams.

*Table 2: Comparison of problem-solving practice, problem-solving decisions[2] and associated template steps.*

| Problem-Solving Practice | Problem-Solving Decisions | Template Steps |
|---|---|---|
| Problem Definition | • Decide what the goals for this problem are | N/A |
| Problem Framing | • Decide what are important underlying features<br>• Decide which predictive frameworks are relevant<br>• Decide what information is relevant/important<br>• Decide what are related problems seen before<br>• Decide how predictive frameworks need to be modified to fit this situation | • Visual Representation<br>• Relevant Concepts<br>• Similar Problems<br>• Assumptions and Simplifications<br>• Information Needed |

| Problem Decomposition and Planning | <ul><li>Decide how to decompose problem into more tractable sub-problems</li><li>Decide what to prioritize</li><li>Decide what information is needed to solve the problem</li><li>Decide on specific plan for getting additional info.</li></ul> | <ul><li>Solution Plan</li><li>Rough Estimate</li></ul> |
|---|---|---|
| Data Interpretation | <ul><li>Decide what calculations are needed</li><li>Decide whether newly obtained information matches expectations</li></ul> | <ul><li>Execution</li><li>Compare to Estimate</li></ul> |
| Reflection on Solution | <ul><li>Decide whether assumptions still make sense</li><li>Decide whether additional information is needed</li><li>Decide how well the solution holds</li></ul> | <ul><li>Check Limits</li><li>Check Units</li><li>Getting Unstuck</li></ul> |

The template was introduced to students in the first week of the course (see Fig. 1) Students were shown an example solution which followed the steps in the template but was flawed. They then worked together in small groups to critique the solution and suggest ways to improve it. It was then explained to the students that this template mirrors the problem-solving process of experts, and that they would use this template to solve problems in class and on their homework assignments. By seeing and improving a flawed example, they could learn what the features of a good, templated solution were, e.g., a detailed and actionable plan. Typically, 3 of the 5 homework problems each week required students to write their solutions on the problem-solving template. Students were graded not just on the accuracy of their final solutions, but also partly on their problem-solving process in following the template steps (e.g., did they make reasonable assumptions?). The template was not required to be used on in-class quizzes, but its use was encouraged to help guide students in timed situations. The in-class worksheets would also follow the steps of the template in guiding students to solve a complex problem using new material. Worksheets would feature worked examples of the template (similar to how it was introduced) and contrasting cases of the template. For example, instead of coming up with their own plan early in the course, students might be shown two contrasting plans and asked to select which one is better. Research has shown that using contrasting cases in this way can help students be more reflective and better evaluate their own learning[11].

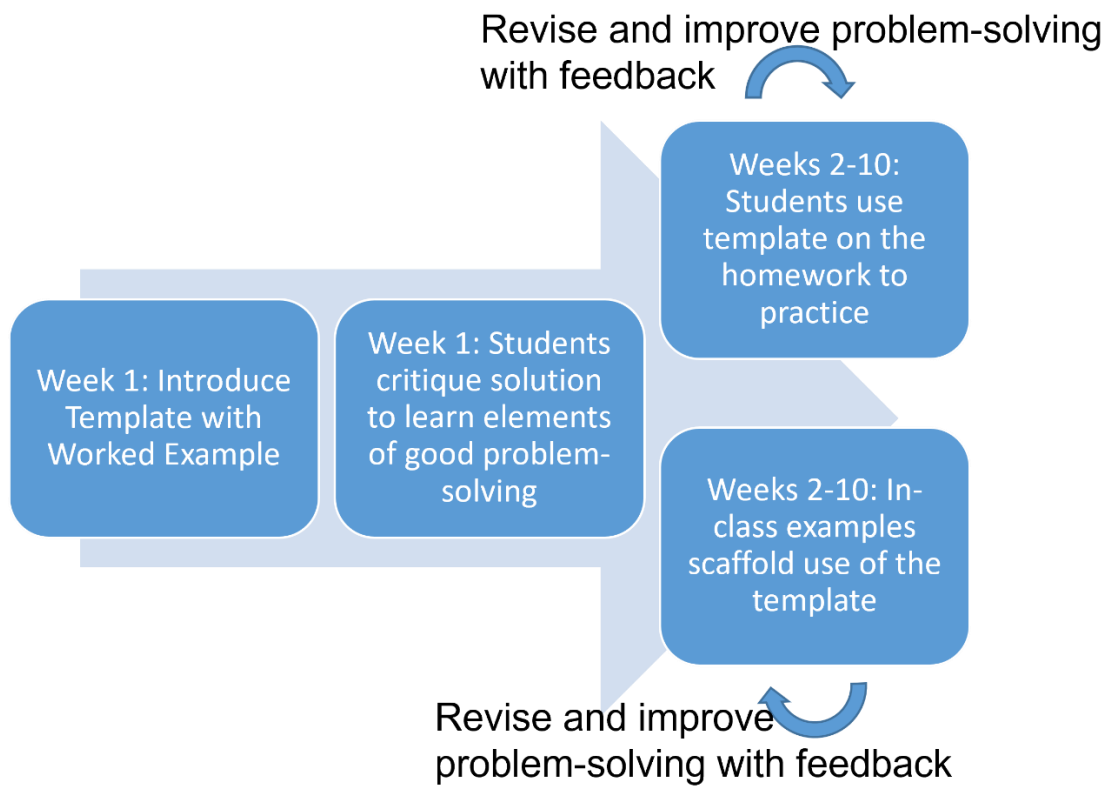

*Figure 1: Graphic illustrating how template was introduced and used in the course.*

**Details of Course Implementation:**

*Student population and selection*

We implemented a careful selection process for our new version of physics 1. Students who wanted to enroll were required to take a physics and math diagnostic test and fill out an online application. The application was focused on the details of their high school physics and math preparation. Students with the lowest levels of high school preparation (measured by diagnostic scores and prior coursework) were given preference for enrollment. Because of the correlation between demographics and the quality of high school preparation, the class was 55% URM, 63% female, and 42% first-generation students – far more diverse than typical introductory physics courses at our institution. Note that unlike many cocurricular support programs (e.g., the Biology Scholars Program[12]), we did not select students based on grit, determination, or other social-psychological factors that may have made them more likely to succeed despite their incoming preparation. The total enrollment in the course was 96 students across two sections. There were 3 instructors of record (each with well-defined but limited roles), 6 graduate teaching assistants (some were only involved in developing new materials), and 11 undergraduate part-time teaching assistants.

*Topic coverage and order.*

*Table 3: List of topics and course schedule for new course.*

| Topic | Description | Week in Term |
|---|---|---|

| Static Forces and Vectors | Newton's 1st + 3rd Law, Vector Algebra, Static Friction | 1 |
|---|---|---|
| Static Torques | Torque in 1D and 2D, $\tau = rF\sin\theta$ | 2 |
| Static Equilibrium | Combined force and torque balances for solving statics problems | 3-4 |
| Conservation of Energy | Gravitational Potential Energy, Kinetic Energy, Spring Potential Energy, Chemical Potential Energy | 5 |
| Conservation of Energy + Work | Work done by friction/drag and how it changes energy balances | 6 |
| Applications of Energy | Applying energy to power plants, batteries, agriculture, etc. | 7-8 |
| Dynamics | 1D Kinematics and problem-solving with constant forces | 9-10 |

The course was organized differently from a standard physics 1 course. We started with static forces and force balances to introduce the fundamental quantity that underlies the physics 1 course and provide practice with vectors. We then introduced torque in the context of statics. We did this because, we realized torque is essential for doing nearly any real statics problem, and statics is essential for understanding force and how it is used in solving real-world problems. We included friction in the statics unit for similar reasons – it plays a role in most realistic statics problems. Note that we introduced all new content in weeks 1-2 and had weeks 3-4 for practice integrating that content to solve problems.

The second part of the course dealt with conservation of energy. We gave it more attention than is typical in physics 1, because our analysis indicated that it played a larger role than other physics 1 material in the courses and problem contexts students would likely encounter in the future. We started with basic conversions between gravitational potential energy and kinetic energy, but then also introduced spring potential energy and chemical potential energy early on. We then introduced work and how to calculate changes in energy from non-conservative forces. Like the statics unit, we had two weeks in which new material was introduced and then two weeks to practice using that new material for problem-solving. The applications of energy also went far beyond what is typical in physics 1, with students considering batteries and solar power, wind power, water tables and crop energy usage, power plant processes and efficiency, etc.

The final unit of the course was dynamics, including 1D kinematics. We also implicitly introduced the ideas of impulse and change in momentum, but just as alternative forms of Newton's second law, without introducing new jargon. We had one week where new information was introduced, and then one week to practice using that information to solve problems.

*Course elements*

The course met 3 times per week for 75 minutes at a time, with an additional 75-minutes per week for students to work on homework problems with TAs present. Design and implementation of in-class activities used worksheets following standard good practices as described in Ref. 5. The worksheets had an overall problem, with several parts students completed that followed the template and required them to produce figures, calculations, and significant text. Students worked in groups of 3 or 4 to complete the worksheets, but each student was responsible for completing and turning in their individual worksheet. They were graded on completion, not quality.

Class followed cycles of students working in small groups for 10-15 minutes with the instructor and TAs circulating and helping, followed by whole class feedback and questions, and then returning to small group work. Typically, during the whole class feedback, students would be given multiple choice questions to answer individually covering parts of the solution. These monitored the understanding of the entire class, as well as providing individual accountability for the students. The specific timing for the whole-class feedback was determined by, a) it was observed most of the groups were stuck on the same issue, or b) approximately 2/3 of the groups had completed a section and all had made good progress on it. Sometimes, the last parts of the worksheet were left for students to complete and turn in as homework. Often the physics content needed to solve the worksheet problem was introduced after students had briefly struggled with the problem, in accordance with the "Preparation for future learning" research[13]. Students were also called upon to decide what information was needed and then find factual information needed to solve the problem by making estimates and/or looking up needed information on the internet. The instructors monitored and lightly intervened on group dynamics and very occasionally rearranged the composition of groups. The overall group composition was kept constant, in response to overwhelming preference by the students when polled.

Homework was assigned on a weekly basis, and all homework problems were written by the instructional team. Typically, one homework problem required students to explain the basic concepts they had learned that week to a lay person, a second tested their conceptual understanding of the content (e.g. identifying forces and torques on an interesting object), and the final 3 problems were template-based problems that were complex, required students to seek out unknown information, and required students to decide how to model a real system. Homework assignments were graded by the TAs, and the template problems were at least partially graded according to the template, not just whether students got their final answers correct. At critical junctures throughout the term, we also included bonus assignments that were implemented through *Atomic Assessments* and automatically graded. These were designed to practice basic skills like identifying forces and doing vector algebra. Homework was worth 20% of students' final grades.

Quizzes were frequent and low states. Four quizzes were administered over the 10-week term, each one worth 10% of students' final grades. The quizzes were two questions each, and students had a full 75-minute class period to complete the quizzes. The quiz questions provided all the information needed, so students did not have to spend time looking up or estimate quantities during the quizzes. This was primarily a pragmatic decision in response to student discomfort with having to find information adding to their stress at taking even a low stakes test. All course notes, homework, homework solutions, worksheets, and internet access were allowed during the quizzes. The only condition was that students not work together on the quizzes. After the quizzes were graded, students were allowed to do quiz corrections to receive 50% of their lost points back. The quiz corrections required students to explicitly (1) identify what they did wrong and (2) identify how they could avoid that mistake in the future. Finally, the lowest quiz grade was dropped.

The final exam was intended to be 9 open-ended questions, 5 of which required the use of the template to solve. It was to be worth 20% of students' final grades, but due to the appearance of COVID-19 at the very end of the term, final exams were cancelled for that term.

The issue of a textbook was challenging for this course. First, we could not find any textbooks that follow the same order of topics and introduce the topics in ways that we did. Second and most important, textbook problems and examples are typically so far removed from reality that their solutions are meaningless as guidance in this course. In lieu of a textbook, we referred students to sections of the Openstax online text and relevant and good Wikipedia articles on selected topics. We also wrote several one-page summaries of important topics (forces, vectors, torques, energy) for students to read prior to class on days when those topics were introduced. Most class periods required no preparation from the students, as so much time was actually spent in class. In retrospect, it was probably more than optimum, and it would be better to have students spent somewhat less time in class and more time working on homework and preparing for class.

**Evaluating Course Effectiveness:**

*Issues of Physics Learning:*

First, we wanted to ensure that students are learning the physics they need to be successful in their future majors and careers. To measure basic understanding of physics, we administered the same Physics Diagnostic to students at the end of the course as in the beginning. This involved doing traditional intro physics problems. Students' scores on the diagnostic increased on average by 20%, 1.1 standard deviations. Students significantly improved on all sub-scores of the diagnostics (even after controlling for multiple comparisons). This suggests that the students successfully learned the basic physics content. This was still slightly below the typical incoming score of the well-prepared students, but this was a 10-week course, and those well-prepared students typically had a year or more of good physics instruction.

*Table 4: P-values for sub-scores on diagnostic (Authors, 2021c) on which students improved.*

| Topic | Increase |
| --- | --- |
| Vectors | 13%, $p = 0.0006$ |
| Math/Trig | 22%, $p = 0.0004$ |
| Equation Reading | 28%, $p < 0.0001$ |
| Force | 16%, $p = 0.0003$ |
| Energy | 28% $p < 0.0001$ |
| Newton's 3rd Law | 20%, $p < 0.0001$ |
| Torque | 27%, $p = 0.001$ |
| **Total** | Change = 20%, $d = 1.1$ stand. dev., $p < 0.0001$ |

Second, we wanted to see if students gained the real-world problem-solving skills that we were attempting to teach. To evaluate this, after the completion of the course we carefully scored the students' problem-solving skills according to the template on three homework problems: one from the first week of the term, one from the fourth week, and one from the ninth week. We found that students' overall problem-solving scores increased steadily from 45% to 55% during the course ($p < 0.001$). Indeed, students showed substantial increases in their ability to plan their solution (55% to 82%, $p < 0.001$) and their ability to follow their proposed plans (60% to 75%, $p < 0.001$). Interestingly, there was a decrease in their problem-representation at the last time point (From 70% to about 48%, $p < 0.001$), as students' visual representation of the problem became less detailed. This may be an indication that students were better able to conceptualize the problems on the later material, and so they did not rely on the external visual representations as heavily.

We also converted problem-solving scores to z-scores for each week, effectively ranking the students at each time point. We found an increase of 0.26 standard deviations from Week 1 to Week 9 ($p = 0.054$), meaning that on average, a student's z-score (ranking in the class) improved by 0.26 over the course of the term. This suggests that the lowest performing students were improving more relative to the rest of the class. Finally, we note that there was a small, steady decrease in the TA-assigned scores over time, but the correlation between TA scores and problem-solving scores substantially increased (from $r = 0.286$, $p < 0.01$ to $r = 0.689$, $p < 0.0001$). We knew the quality of the TA grading was an issue, but this suggests that the TAs were getting better at grading students according to the template over time, likely due to our nagging.

*Interviews with former students:*

In addition to the quantitative analysis, we interviewed former students to find out what aspects of the course they thought worked and what did not. Quotes from students suggest that, while there were areas for improvement in the course, we were successful in our mission to teach students how to solve real physics problems:

> *"Even if I didn't learn every single concept that a 41[the standard physics 1 course] student might have learned, I feel like I came out of the class better equipped than a 41 student to tackle complex problems in any field."*

> *"I've heard my other friends making comments about like 'The one thing I took from [this course] was that no matter how weird a problem is, I'm not afraid to approach it anymore.'"*

The interviews also suggested that the template was useful not only in our course, but in future courses:

> *"I found it especially helpful at times when I was just lost in a problem and didn't really have any direction. And so I remember one time I was really lost in a problem and because of the template it really encouraged me to go back to the question stem and just parsing out the important parts from the question and question stem. The template really forces you to do that and that really allowed me to solve the problem."*

> *"At first it was annoying, but I find myself using it in other engineering classes."*

From a social-psychological perspective, the interviews and surveys indicated that students also felt more confident in physics, well-prepared for Physics 2 and Engineering Statics, felt that the course positively influenced their choice to stay in STEM, and felt that the class was inclusive.

*Issues of Equity and Access:*

One of our primary goals was to ensure that students could succeed in physics 1, regardless of their prior physics preparation. Table 5 suggests that we were successful in that goal, as there is no correlation between FMCE score, or their score on our physics diagnostic test and a students' total quiz grade, in stark contrast to what we had previously seen in the standard physics 1 course at our institution. There is a small correlation between SAT/ACT math score and quiz grade, indicating the importance of trigonometry and algebra skills for success in the course. Future implementations of this course should include more opportunities for students to practice trigonometry and algebra skills early on in the course, to avoid these being barriers to future success in the class. Finally, we note a strong correlation between diagnostic post-score and quiz grade. This suggests that the course does indeed cover much of the same content as the physics diagnostic, so the lack of correlation with prior physics preparation is not simply because the course covers different material.

Table 5: Pearson correlation between quiz grade and various measures of incoming preparation. * p < 0.05, *** p < 0.001.

|  | Correlation with Quiz Grade |
|---|---|
| **SAT/ACT Math Score** | r = 0.288* |
| **FMCE Pre-Score** | r = 0.095 |
| **Diagnostic Pre-Score** | r = 0.137 |
| **Diagnostic Post-Score** | r = 0.548*** |

**Conclusion:**

In this article, we described the design and implementation of an introductory Mechanics course that enabled students to learn critical physics content and problem-solving skills. The problems solved in this course were much closer to the kinds of problems students will need to solve in their future careers as scientists and engineers. In addition, students' success in the class was independent of their prior physics preparation. Designing courses to be independent of incoming preparation while supporting future academic success is important for improving equity in STEM education and increasing representation of historically marginalized groups in physics. We hope that this course will provide a useful model for other institutions and are happy to share the materials.